# Determining On-Axis Crystal Thickness with Quantitative Position-Averaged Incoherent Bright-Field Signal in an Aberration-corrected STEM

Huolin L. Xin,*[1] Ye Zhu,** and David A. Muller**,***

* Department of Physics, Cornell University, Ithaca, NY 14853

** School of Applied and Engineering Physics, Cornell University, Ithaca, NY 14853

*** Kavli Institute at Cornell for Nanoscale Science, Cornell University, Ithaca, NY 14853

## Abstract

An accurate determination of specimen thickness is essential for quantitative analytical electron microscopy. Here we demonstrate that a position-averaged incoherent bright-field signal recorded on an absolute scale can be used to determine the thickness of on-axis crystals with a precision of $\pm 1.6$ nm. This method measures both the crystalline and the non-crystalline parts (surface amorphous layers) of the sample. However, it avoids the systematic error resulting from surface plasmons contributions to the inelastic mean free path thickness estimated by electron energy loss spectroscopy.

[1] Email: hx35@cornell.edu; current address: 1 Cyclotron Road Mail Stop 66-227, LBNL, Berkeley, CA 94720

## Introduction

Transmission electron microscopy provides projected images of three-dimensional samples. Details of the specimen thickness are lost in any given image or spectroscopic map. While the full shape of the sample can be reconstructed from electron tomography, the method is time consuming and often dose intensive. For quantitative analytical work, faster and simpler methods of thickness estimation are desirable. Often, simple inspection of an annular dark-field scanning transmission electron microscope (ADF-STEM) image, electron energy loss spectroscopic (EELS) map or x-ray map can provide useful qualitative information (Bosman, et al., 2007; Botton, et al., 2010; D'Alfonso, et al., 2010; Kourkoutis, et al., 2010; Muller, 2009; Muller, et al., 2008; Xin, et al., 2011). However, when properly calibrated and corrected for thickness-dependent artifacts, it is possible to extract quantitative, atomic-scale information such as composition (Crewe, et al., 1970; Krivanek, et al., 2010; Radmilovic, et al., 2011; Voyles, et al., 2002), lattice structures and their distortions (Borisevich, et al., 2010; Fitting, et al., 2006; Mkhoyan, et al., 2006), and the number of atoms in atomic columns (LeBeau, et al., 2010b; Li, et al., 2008; Van Aert, et al., 2011). Although ADF images are often qualitatively interpretable, the intensity and contrast of ADF-STEM images are not linear or simple functions of specimen thickness (Fitting, et al., 2006; Hillyard, et al., 1993; Hillyard & Silcox, 1993; Kirkland, 1998; Kirkland, et al., 1987; Klenov, et al., 2007; Kourkoutis, et al., 2011; LeBeau, et al., 2009; LeBeau & Stemmer, 2008; Lupini, et al., 2009; Mittal & Andre Mkhoyan, 2011; Voyles, et al., 2004; Xin, et al., 2008; Xin & Muller, 2010). In order to quantify the atomic-scale information in projection, a multislice frozen-phonon simulation of the sample at the correct thickness is required. Therefore, an accurate determination of specimen thickness is a prerequisite for quantitative scanning transmission electron microscopy. In addition, an accurate thickness calibration is also needed for quantitative spectroscopy such as spectral deconvolution in electron energy loss spectroscopy (EELS) and absorption corrections in energy-dispersive X-ray spectroscopy. In this paper, we evaluate the use of a quantitative position-averaged incoherent bright-field (QPA-IBF) signal to measure the thickness of on-axis crystalline samples. This technique, requiring no change to the optical setup of image acquisition, is well suited for any high-resolution aberration-corrected STEM imaging of on-axis crystals.

For on-axis crystalline samples, there are at least three popular methods of determining the crystal thickness in a STEM. First, the thickness can be determined by measuring $t/\lambda$, the thickness as a fraction of the inelastic mean free path, using an electron energy loss spectrometer (EELS) (Egerton, 2011). ($t$ is the thickness and $\lambda$ is the inelastic mean free path). Second, thickness of the crystalline portion of the sample can be determined by position-averaged convergent beam electron diffraction (PA-CBED) patterns (Kourkoutis, et al., 2011; LeBeau, et al., 2010a; Loane, et al., 1991). By matching the thickness fringes in the recorded PA-CBED pattern with a dynamical simulation, the thickness can be determined (Kourkoutis, et al., 2011; LeBeau, et al., 2010a). Third, measuring the annular dark-field (ADF) signal on an absolute scale with a spatially calibrated detector also enables a standardless estimate of specimen thickness (LeBeau, et al., 2008; LeBeau, et al., 2010b). In general, however, each of these thickness-determination methods has known systematic errors and weakness. For example, EELS estimates encounter difficulties in modeling surface plasmons and Cherenkov radiation that lead to systematic offsets in apparent thickness (Mkhoyan, et al., 2007); CBED and diffraction measurements neglect amorphous surface layers; ADF signals can encounter contrast reversals in thick specimens--such as those that are involved in typical semiconductor devices (Ercius, et al., 2009; Ercius, et al., 2006) and the non-uniform response across the detector to the scattered signal makes alignment and reproducibility of the results very challenging. Therefore, being able to obtain more than one independent thickness measurement can help to identify the sources of errors and potentially correct for them. Here we evaluate the use of position-average incoherent bright-field signal (PA-IBF) as a means of thickness determination. We found that it is insensitive to defocus, calibration of the detector is relatively simple, and it does not require detailed knowledge of the probe. In addition, it is sensitive to both crystalline and amorphous portions of the sample. While both position-averaged ADF (PA-ADF) and PA-IBF signals could be used for thin samples, for thicker samples, the PA-IBF is more robust.

The incoherence of the bright-field STEM signal is achieved by collecting both the unscattered bright disk and a large portion of the scattered electrons. The collection semi-angle should be chosen more than 2-3 times the convergence semi-angle of the incident probe and it should also be larger than the third Bragg disks of the zone axis (Hartel, et al., 1996; Howie, 1979). Once these two criteria are met, the bright-field signal is approximately incoherent and it is a monotonically decreasing function of the specimen thickness. The IBF signal in STEM was discussed theoretically by Sheppard (Sheppard & Choudhury, 1977), Rose (Hartel, et al., 1996; Rose, 1975) and Howie (Howie, 1979). It was not widely used because for thin specimens the signal-noise-ratio (SNR) of IBF-STEM images (small negative signals on a full-beam-intensity background) is much lower than that of ADF-STEM images (positive signals on a zero-intensity background) (Ercius, et al., 2009). However, IBF-STEM becomes more useful in thicker specimens because it is free of the contrast reversals exhibited in ADF-STEM due to scattering outside the detector, and can have a better signal-to-noise ratio (Ercius, et al., 2009; Ercius, et al., 2006; Rose & Fertig, 1976). The monotonic response of the IBF signal as a function of thickness makes it especially useful in the tomographic imaging of thick specimens where a reliable reconstruction requires the signal to be monotonic (Ercius, et al., 2009; Ercius, et al., 2006).

As the IBF signal is a monotonically decreasing function of thickness, we should be able to uniquely determine the thickness if we can measure the IBF signal on an absolute scale (i.e. IBF signal normalized by the incident beam intensity). The quantification of the signal on an absolute basis enables the matching between experiments and simulations. If we further remove the position dependency of the signal (the dependency of the beam's relative position to the crystal unit cell), in principle we should be able to eliminate any variation in the signal from effects such as the incoherent source size, the aberrations and the defocus value. This will facilitate a much easier matching between the simulation and the experiment. After meeting these conditions, we use the quantitative position-averaged incoherent bright field signal (QPA-IBF), which we define as the incoherent bright field signal averaged over a repeating unit cell normalized by the incident beam intensity (Figure 1). The thickness can be determined once the QPA-IBF signal vs. thickness function is calculated using the multislice frozen-phonon simulation.

In this paper, we will demonstrate theoretically that the QPA-IBF method is a monotonically decreasing function of thickness and it is strictly independent of defocus and other aberrations in thin specimens. The response is still robust for thicker samples. We experimentally demonstrate the method in an aberration-corrected STEM using a DyScO$_3$ (DSO) single crystal. We show that he QPA-IBF determined thickness is linear with the $t/\lambda$ determined by EELS. We also compare the QPA-CBED method with the PA-CBED method on a SrTiO$_3$ (STO) crystal using an uncorrected STEM to show that the QPA-CBED method measures both the crystalline and amorphous parts of the sample.

# Materials and Methods

## Experimental Setup

In an aberration-corrected Nion UltraSTEM, we used a DSO single crystal specimen oriented along the [101] zone axis for a test of the QPA-IBF method. The specimen was prepared by focused ion beam lift-out. The final surface was finished with 2 kV milling to reduce the surface damage layer. The aberration-corrected Nion UltraSTEM was operated at 100 keV with a convergence semi-angle of 31.8 ±0.1 mrad (beam current ~100 pA). The IBF signal and the electron energy loss spectra were recorded using a Gatan Enfina spectrometer. The IBF and the EELS collection semi-angle (79.8 ±0.1 mrad) were defined by the inner boundary of the ADF detector. We used the quadruple lens Q3 in the spectrometer to demagnify the energy-loss signal in the non-dispersing direction to ensure that every electron that went through the inner hole of the ADF detector can be recorded on the spectrometer CCD (Figure 2a). A dispersion of 1.1 eV/channel was used collecting electron energy loss spectra from -110 eV to 1360 eV. The large energy range allows most inelastically scattered electrons to be recorded. Electrons that fall outside this energy range account for less than 0.2% of the incident beam, and less than 0.3% of the collected signal in the thickness range (20-85nm). Therefore, the sum of the intensity of each spectrum is essentially the IBF signal (Figure 2b). A 64 ✕ 64 pixel EELS map was recorded over a 20 Å × 20 Å area which is approximately 2.5 × 2.5 unit cells (or 5 × 5 pseudo-cubic unit cells) (Figure 2c). The ultrafast electrostatic beam blanker was used to enable an integration interval of 0.1 msec/pixel to prevent the saturation of the detector. (The blanker is synchronized with the spectrometer to blank the beam during the read-out). The IBF image is built by integrating the spectroscopic map over the energy axis. Typical for a cold field-emission instrument, the beam current slowly decays over the course of several hours. Hence the beam current was reinstated every hour by passing a high current through the field-emission tip to evaporate off the adsorbates. Therefore, following each spectroscopic map on the sample, the incident beam intensity was measured by the spectrometer by recording the zero-loss peak in vacuum. The quantitative IBF image is constructed by dividing the IBF image by the incident beam intensity. The QPA-IBF signal is the average of the IBF image over an integer of unit cells. A properly calibrated bright field detector, such as that made from a scintillator and photomultiplier tube, could also be used to collect the IBF signal, provided the detection electronics could accommodate the large dynamic range of the signal (Ercius, et al., 2006). This would have the disadvantage of blocking the signal needed for core-loss EELS, but might be very convenient for x-ray mapping. Using the signal integrated on the spectrometer is compatible with recording the full energy loss range needed for quantitative EELS.

The QPA-IBF method, similar to the PA-CBED method, is dependent on dynamical simulations. Both the Bloch wave simulation and the multislice frozen phonon simulation give the correct thickness fringes in the calculation of on-axis PA-CBED patterns. However we find that the Bloch wave method introduces systematic errors in simulating the IBF signal for all but the thinnest samples due to an insufficient treatment of thermal diffuse scattering, similar to that encountered in ADF image simulation (LeBeau, et al., 2008). Unlike PA-CBED which only measures the thickness of the crystalline portion of a sample, the QPA-IBF method also takes into account the amorphous layers on the top and bottom sample surfaces. To test the difference between the two methods, we carried out a comparison of both methods on a common perovskite crystal—$SrTiO_3$—using a Tecnai F20 (200 kV, Schottky field-emission gun, S/TEM). High-quality PA-CBED patterns can be recorded on a Gatan GIF CCD in this instrument allowing the crystalline thickness to be determined. The difference between the two methods should be approximately equal to the sum of the amorphous layers on the top and bottom surfaces of the sample. To confirm that the extracted thickness of the amorphous layers is physically meaningful, we used two separate $SrTiO_3$ (STO) wedge polished and ion-milled samples. One was an as-milled sample. The other sample was HF-dipped prior to imaging.

## Theoretical Calculations

The QPA-IBF signal was calculated by a multislice simulation with the frozen phonon model. Calculations were performed with the autostem program coded by Earl Kirkland (Kirkland, 2010). It can be downloaded online at http://people.ccmr.cornell.edu/~kirkland/. A 23x23 angstrom$^2$ super cell for DSO and a 20x20 angstrom$^2$ super cell for STO with a 1024x1024 pixel$^2$ sampling were used in the simulation. A 32x32-pixel$^2$ unit-cell image was simulated for each defocus value. The QPA-IBF signal is the average of the simulated unit-cell image. The crystal structure and the Debye-Waller factor of DSO and STO were obtained from the literature (Liferovich & Mitchell, 2004; Peng, 2005). Eight Einstein phonon configurations were used for each simulation.

# Results and Discussions

## Theory of Position-Averaged Incoherent Bright Field Signal

*Thin specimen regime*

In the single scattering regime, the IBF signal is the complement of an ideal ADF signal (i.e. the annular dark-field signal with the outer collection angle going to infinity and with the inner angle being equal to that of the maximum collection angle of IBF), assuming no electrons come to rest in the sample – a good approximation for samples less than a few tens of microns thick. As along as the maximum IBF collection angle is larger than the incident beam convergence angle, the IBF-STEM image can be formulated as

$$I(\mathbf{r}) = \int B(\mathbf{k}) \left| \int A(\mathbf{k}')V(\mathbf{k},\mathbf{k}') \exp(-i\chi(\mathbf{k}') - 2\pi i \mathbf{k}'\mathbf{r}) \, d\mathbf{k}' \right|^2 d\mathbf{k} \qquad (1)$$

where $\mathbf{r}$ is the relative position of the incident beam with the origin, $\mathbf{k}'$ is the transverse wave vector of the incident plane waves, and $\mathbf{k}$ is the transverse wave vector of the outgoing plane waves. $A(\mathbf{k}')$ is the aperture function which is defined as

$$A(\mathbf{k}') = \begin{cases} \dfrac{1}{\sqrt{\pi}\alpha_{max}/\lambda}, & |\mathbf{k}'| \leq \alpha_{max}/\lambda \\ 0, & otherwise \end{cases} \qquad (2)$$

Where $\alpha_{max}$ is the convergence semi-angle of incident probe and $\lambda$ is the wavelength of the electrons. $B(\mathbf{k})$ is the detector function:

$$B(\mathbf{k}) = \begin{cases} 1, & |\mathbf{k}'| < \beta_{max}/\lambda \\ 0, & otherwise \end{cases} \qquad (3)$$

Where $\beta_{max}$ is the IBF collection semi-angle. For ideal ADF imaging, $B_{ADF}(k) = 1 - B(k)$.

$\chi(k')$ is the phase shift due to geometrical aberrations and defocus. If only taking the spherical aberrations into account, the expression for $\chi$ up to the seventh order is

$$\chi(k') = \frac{2\pi}{\lambda}\left[\frac{C_3\lambda^4 k'^4}{4} + \frac{C_5\lambda^6 k'^6}{6} + \frac{C_7\lambda^8 k'^8}{8} + O(k'^{10})\right] - \pi df \lambda^2 k'^2 \quad (4)$$

Where $C_3$, $C_5$ and $C_7$ are the third, fifth and seventh order spherical aberration coefficients respectively and $df$ is defocus. $V(k, k')$ is the elastic scattering transition potential. For a single crystal of a single element, it can be written as

$$V(k, k') = \sum_G f(k' - k)\delta(k' - k - G) \quad (5)$$

where the $f(k' - k)$ is the scattering factor of a single atom; (the sum over different atoms is not expressed here without loss of generality;) $G$ is the Bravais lattice vector. In a 1-D crystal, $G = \frac{n}{a}e_x$ $(n = \cdots, -2, -1, 0, 1, 2, \cdots)$, where $a$ is the lattice constant and $e_x$ is the unit vector in the $x$ direction.

If we substitute (5) into (1) and use the Kronecker-delta relation $k' = k + G$, the equation reads

$$I(r) = \int B(k)\left|\sum_G A(k+G)f(G)\exp(-i\chi(k+G) - 2\pi i(k+G)r)\right|^2 dk$$

$$= \int B(k)dk \sum_G A(k+G)f(G)\exp(-i\chi(k+G) - 2\pi i(k+G)r) \quad (6)$$

$$\times \sum_{G'} A(k+G')f(G')\exp(i\chi(k+G') + 2\pi i(k+G')r)$$

Simplify equation (6), it reads

$$I(r) = \sum_{G,G'} f(G)f(G')\exp(-2\pi i(G - G')r)$$

$$\times \int B(k)A(k+G)A(k+G')\exp(-i(\chi(k+G) - \chi(k+G')))dk \quad (7)$$

Equation (7) is the formulation of the IBF image of a thin crystal. Due to the explicit appearance of χ($k$) in the equation, the contrast of the IBF image is dependent upon the aberrations and defocus of the incident beam. However, if we carry out a position averaging of the image in a unit cell, we can demonstrate the phase shift due to defocus and aberrations can be canceled out. Here, without the loss of the generality, we assume the crystal is a 1-D crystal. The position-averaged IBF signal, therefore, is

$$\int_0^a I(r)dr/a = \frac{1}{a}\sum_{G,G'} f(G)f(G') \int_0^a \exp(-2\pi i(G-G')r)dr \int B(k)A(k+G)A(k+G')\exp(-i(\chi(k+G)-\chi(k+G')))dk \tag{8}$$

Due to integration over the sinusoidal term $\int_0^a \exp(-2\pi i(G-G')r)\,dr$, the only nonzero contributions in the summation over $G$ and $G'$ are those that satisfies $G = G'$. Therefore, the phase shift χ($k+G$) − χ($k+G'$) cancels out in equation (7). We have

$$\text{QPA-IBF} = \sum_G f^2(G) \int B(k)A(k+G)A(k+G)dk \tag{9}$$

As shown by equation (9), the QPA-IBF signal is independent of aberrations and defocus of the incident beam in a thin specimen where multiple scattering is ignorable. Similarly, by replacing $B(k)$ by $B_{ADF}(k)$ in equation (9) we would obtain the QPA-ADF signal as 1 minus the QPA-IBF signal, which again would be independent of aberrations and defocus.

*Thick specimen regime*

The above derivation is only correct for thin specimens in the single scattering regime. However, as long as the IBF collection angle is large enough, multiple scattering in a finite thick crystal does not change the conclusion. As shown in Figure 3, in both $SrTiO_3$ and a heavy element containing crystal—$DyScO_3$—the QPA-IBF signal is a monotonic decreasing function and it is nearly independent of defocus. The QPA-ADF signal however can undergo contrast reversals in thick samples as electrons are scattered out beyond the outer angle of the detector (Ercius, et al., 2006).

## Using QPA-IBF to determine on-axis crystal thickness

As shown in the above analytic derivation and the multislice simulations, the position averaging over a unit cell efficiently removes any variation in the signal from effects such as the defocus value. Therefore, once the QPA-IBF signal is measured, the thickness can be determined using the simulated QPA-IBF vs. thickness curve such as the ones shown in Figure 3. To test the precision and accuracy of this method, we compared the QPA-IBF method with both the $t/\lambda$ method and the PA-CBED method.

*QPA-IBF vs. $t/\lambda$ in an aberration-corrected STEM*

As we stated above, the $t/\lambda$ measured by EELS gives a good estimate of the relative thickness of the sample. However, it lacks the ability to provide an absolute thickness measurement due to difficulty in calculating the angular dependent plasmon contributions, especially if the conduction band of the crystal cannot be approximated as a free electron gas. In addition, the $t/\lambda$ method is complicated by surface-plasmon generation. In Si for instance the surface contribution is approximately equal to 2 nm of the bulk when the sample is around 65 nm thick (Mkhoyan, et al., 2007). Errors as large as 10 nm in thickness are not unusual (Egerton, 2011; LeBeau, et al., 2008)

So, if our QPA-IBF method is accurate, we expect the QPA-IBF determined thickness to be linear with $t/\lambda$, and we also expect a positive intercept on the $t/\lambda$ axis due to the generation of surface plasmons. Figure 4 shows the QPA-IBF determined thickness vs. $t/\lambda$. The $R^2$ value of the fitting is 0.9922 indicating the QPA-IBF thickness is nearly linear with $t/\lambda$. The standard deviation of the residuals is ±1.64 nm. This indicates the precision of our QPA-IBF method is around ±1.6 nm (assuming $t/\lambda$ is a much more precise measurement). The fitted linear expression is $t_{QPA-IBF} = (79.8 \pm 1.4) \times t/\lambda - (1.19 \pm 0.85)$ (nm). (The uncertainty bounds here are the 68.2% confidence intervals.) The fitted line intercepts $t/\lambda$ axis on the positive side with statistical significance (p-value = 0.08) as expected. The contribution of the surface plasmons is equal to the bulk contribution of a 1.19nm-thick layer of $DyScO_3$, of similar magnitude to that reported for silicon (Mkhoyan, et al., 2007).

*QPA-IBF vs. PA-CBED*

Compared to the PA-CBED method, the QPA-IBF method not only measures the thickness of the crystalline region of the sample, it also takes into account the amorphous layers on the top and bottom surfaces. To measure the difference in the two methods, we carried out a comparison of the two methods on a single crystal $SrTiO_3$.

As shown in Figure 5(b), the QPA-IBF method consistently gives a thickness measurement higher than that of the PA-CBED method due to the amorphous surface layers. In the as-milled sample, the amorphous layer is on average 9.8 ±1.1 nm (i.e 4.9 nm/surface). In the HF-dipped sample, the amorphous layer is on average 6.4 ±1.1 nm (3.2 nm/surface). Both these amorphous layers are larger than we typically see in freshly prepared samples (these had been reimaged multiple times) and the image contrast was also not as crisp.

*Discussions*

The QPA-IBF method, similar to the PA-CBED method, is dependent on multi-beam dynamical simulations. It should in principle be accurate if the crystal structure, the Debye–Waller factors, the incident semi-angle and collection semi-angle are determined correctly and if the detector response is linear. Also like PA-CBED, the QPA-IBF signal is influenced by the lower-order structure factors, which are potentially affected by the bonding environment. However, the error due to bonding changes in the lower-order structure factors should be smaller for crystals containing heavy elements due to the relatively low ratio of valence electrons to the nuclear charges.

In addition, since this method requires a quantitative matching between the simulation and the experiment, the accuracy of the QPA-IBF determined thickness is highly dependent upon the linearity of the detector on which the IBF signal is recorded. To demonstrate potential artifacts, we also recorded the QPA-IBF signal on our Ronchigram CCD following the IBF map acquisition using the spectrometer. The measurement of the linearity of the Ronchigram CCD shows that the Ronchigram CCD is highly nonlinear prior to saturation (Figure 6a). The response at higher intensity is higher than that extrapolated from the lower-intensity data points. The high-contrast nonlinearity of the Ronchigram CCD results in an underestimate of the QPA-IBF signal and thus an overestimate of the thickness. On the other hand, the spectrometer CCD is nearly linear until close to saturation (Figure 6b). Therefore, checking the linearity of the recording detector is critical for the accuracy of this method.

## Conclusions

In this paper, we have reported a robust method of determining the thickness of on-axis single crystalline samples using a quantitative position-averaged incoherent bright-field (QPA-IBF) signal, suitable for use in both uncorrected and aberration-corrected scanning transmission electron microscopes. By recording the position-averaged IBF signal normalized by the incident beam intensity, the thickness can be determined by matching the QPA-IBF signal with frozen-phonon simulations. The QPA-IBF determined thickness takes into account both the crystalline region and the amorphous layers on the top and bottom surfaces of the sample. It is on average thicker than the PA-CBED determined thickness due to the presence of amorphous surface layers. However, it excludes the systematic error resulting from surface plasmons in the valence EELS $t/\lambda$ method. The acquisition of the QPA-IBF signal does not require any change of the imaging optics. It is well suited for an aberration-corrected STEM without a high-dynamic range imaging CCD for PA-CBED but equipped with an EELS spectrometer, or with a high dynamic range bright-field point detector. The QPA-IBF derived thickness is similar to that from ADF for thin samples, but is more robust in thicker samples. Calibration of the detectors used for QPA-IBF are often easier to perform and less sensitive to small errors in alignment than ADF. In general and given the different systematic errors in each method, using more than one method to determine thickness is desirable, and the QPA-IBF recorded on an EELS system often also contains the information for extracting a $t/\lambda$ thickness measurement as well.

## Acknowledgement

HLX thanks Christian Dwyer and Robert Hovden for proof reading the manuscript and providing insightful input. HLX and YZ thank Lena Fitting Kourkoutis for supplying the as-milled $SrTiO_3$ specimen. HLX was supported by the Energy Materials Center at Cornell (EMC2), an Energy Frontier Research Center funded by the U. S. Department of Energy, Office of Science, Office of Basic Energy Sciences under Award Number DE-SC0001086. YZ was supported by the Army Research Office award W911NF0910415. This work made use of the electron microscopy facility of the Cornell Center for Materials Research (CCMR) with support from the National Science Foundation Materials Research Science and Engineering Centers (MRSEC) program (DMR 1120296) and NSF IMR-0417392.

# Figure Captions

Figure 1. (color online) Schematic showing the quantitative position-averaged incoherent bright field (QPA-IBF) signal. (The drawings are not to scale.)

Figure 2. (color online) The quantitative IBF signal recorded on an Enfina spectrometer. (a) Quadrupole Q3 was adjusted to demagnify the image in the non-dispersive direction such that the whole IBF collection aperture can be recorded on the spectrometer CCD. The zero loss peak shown here has been intentionally defocused in the energy dispersive direction to reveal the bright-field disk and the IBF collection aperture. (b) The experimental EELS spectra. The quantitative IBF signal is defined by the integrated intensity of the through-sample spectrum divided by the integrated intensity of the through-vacuum spectrum. (c) Atomic-resolution quantitative IBF image from $DyScO_3$.

Figure 3. (color online) Simulated QPA-IBF signal from $DyScO_3$ [101] (100 keV, convergence semi-angle 36 mrad, IBF collection angles 0-75 mrad) and $SrTiO_3$ [001] (200 keV, convergence semi-angle 10.1 mrad, IBF collection angles 0-30 mrad). The range of the selected defocii is larger than the depth of focus of the respective optical conditions. The QPA-IBF signal decreases monotonically with thickness and is nearly independent of defocus.

Figure 4. (color online) QPA-IBF determined thickness vs. valence EELS recorded $t/\lambda$ on $DyScO_3$ [101]. The fitted linear expression is $t_{QPA-IBF} = 79.8 \times (t/\lambda - 0.0149) = 79.8 \times t/\lambda - 1.19$ (nm).

Figure 5. (color online) Comparison of PA-CBED and QPA-IBF methods. (a) The determination of crystalline thickness by matching the PA-CBED with the incoherent CBED simulation. (b) QPA-IBF thickness vs. PA-CBED thickness. The QPA-IBF method gives a thickness measurement higher than of the PA-CBED method due to the presence of amorphous surface layers. In the as-milled sample, the amorphous layer is on average 9.8±1.1 nm (i.e. 4.9 nm/surface). In the HF-dipped sample, the amorphous layer is on average 6.4±1.1 nm (3.2 nm/surface). The HF-dipped sample and the as-milled sample are from two separate STO specimen.

Figure 6. (color online) The comparison of the Ronchigram CCD and the spectrometer CCD. (a) The measurement of the linearity of the Ronchigram CCD. The Ronchigram CCD is highly nonlinear prior to saturation. (b) The measurement of the linearity of the spectrometer CCD. The spectrometer CCD is very linear until saturation.

Figures and Captions

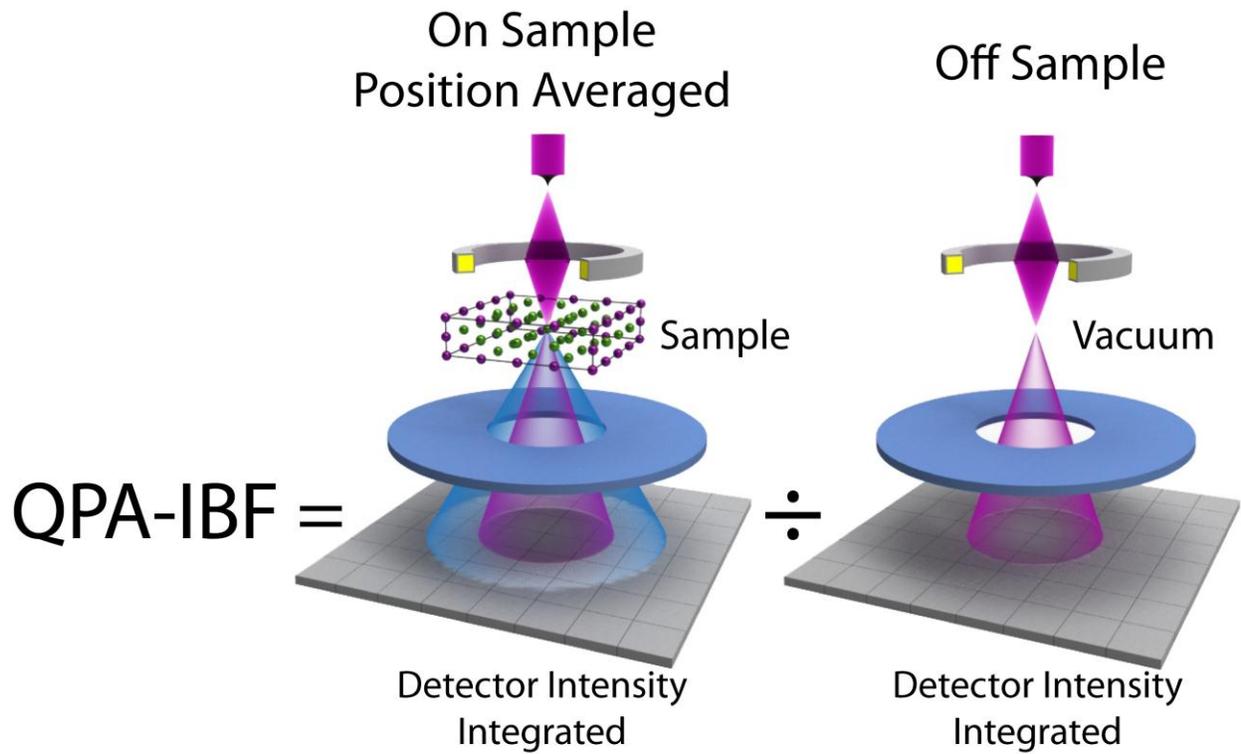

Figure 1. (color online) Schematic showing the quantitative position-averaged incoherent bright field (QPA-IBF) signal. (The drawings are not to scale.)

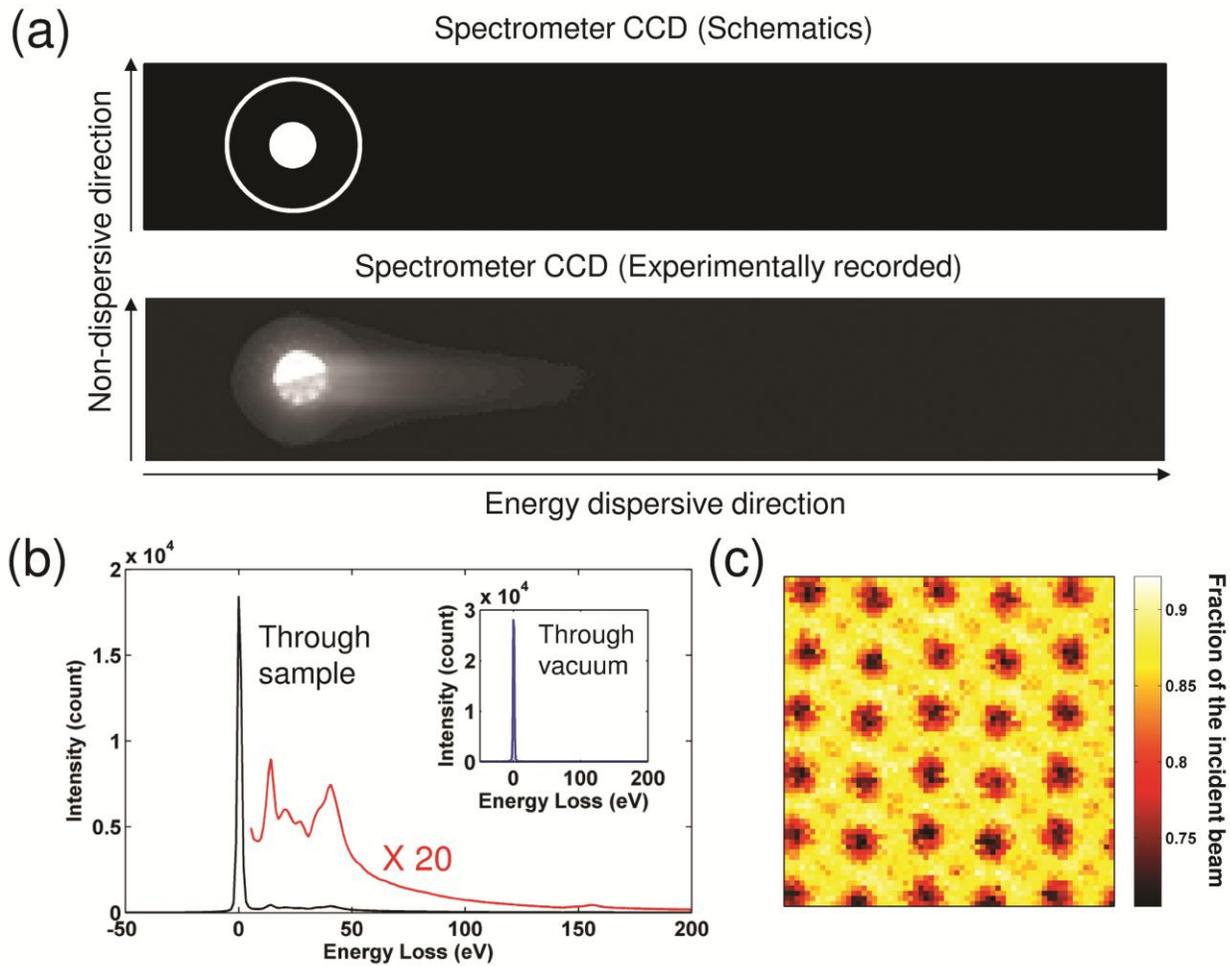

Figure 2. (color online) The quantitative IBF signal recorded on an Enfina spectrometer. (a) Quadrupole Q3 was adjusted to demagnify the image in the non-dispersive direction such that the whole IBF collection aperture can be recorded on the spectrometer CCD. The zero loss peak shown here has been intentionally defocused in the energy dispersive direction to reveal the bright-field disk and the IBF collection aperture. (b) The experimental EELS spectra. The quantitative IBF signal is defined by the integrated intensity of the through-sample spectrum divided by the integrated intensity of the through-vacuum spectrum. (c) Atomic-resolution quantitative IBF image from $DyScO_3$.

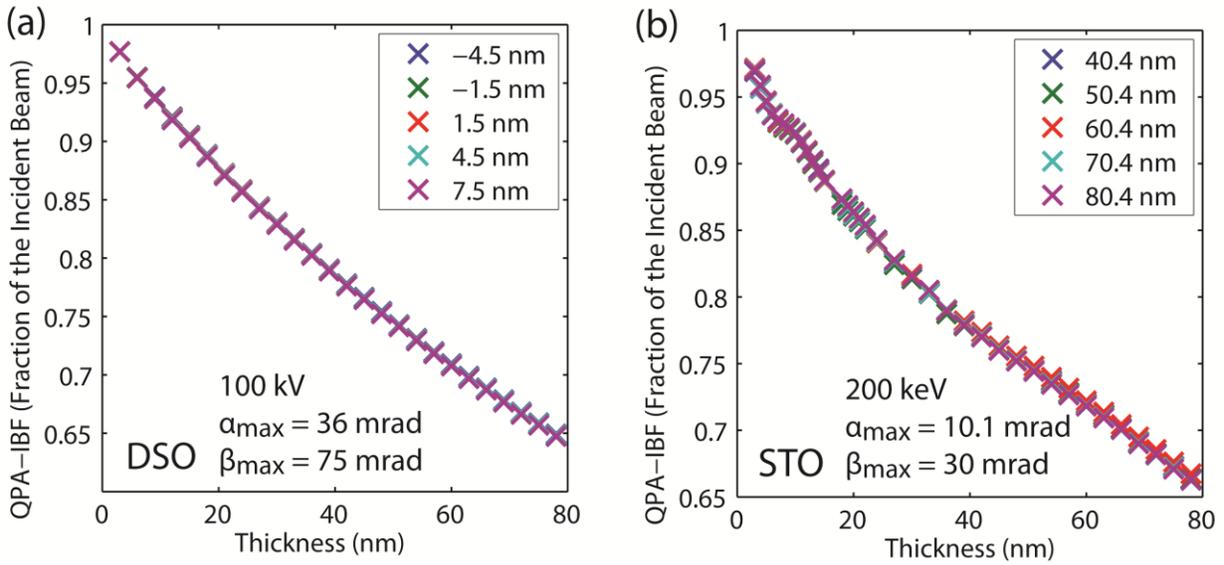

Figure 3. (color online) Simulated QPA-IBF signal from DyScO$_3$ [101] (100 keV, convergence semi-angle 36 mrad, IBF collection angles 0-75 mrad) and SrTiO$_3$ [001] (200 keV, convergence semi-angle 10.1 mrad, IBF collection angles 0-30 mrad). The range of the selected defocii is larger than the depth of focus of the respective optical conditions. The QPA-IBF signal decreases monotonically with thickness and is nearly independent of defocus.

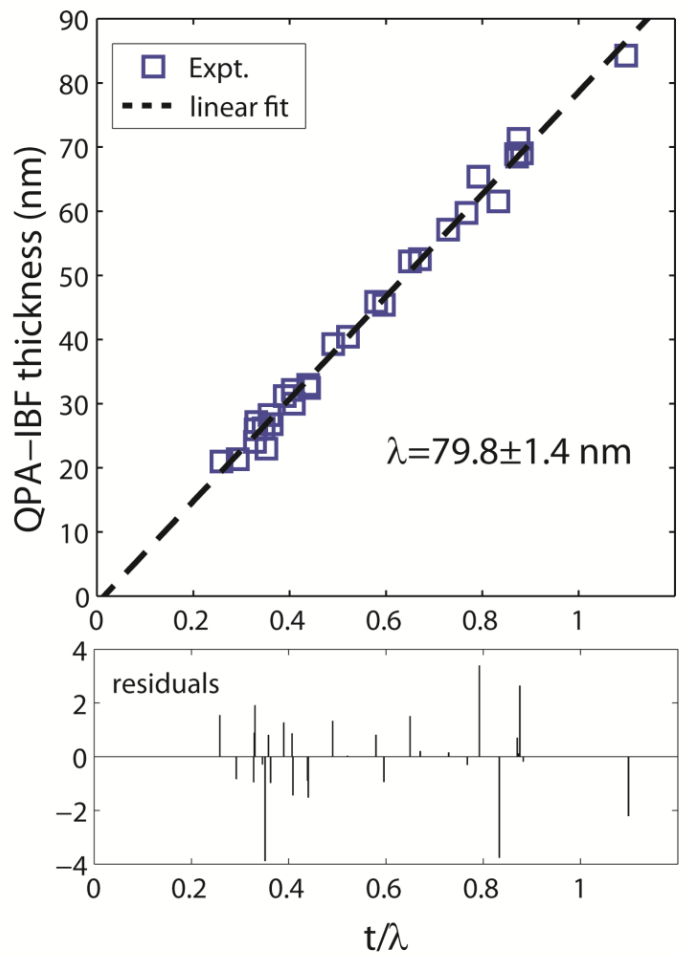

Figure 4. (color online) QPA-IBF determined thickness vs. valence EELS recorded $t/\lambda$ on DyScO$_3$ [101]. The fitted linear expression is $t_{QPA-IBF} = 79.8 \times (t/\lambda - 0.0149) = 79.8 \times t/\lambda - 1.19$ (nm).

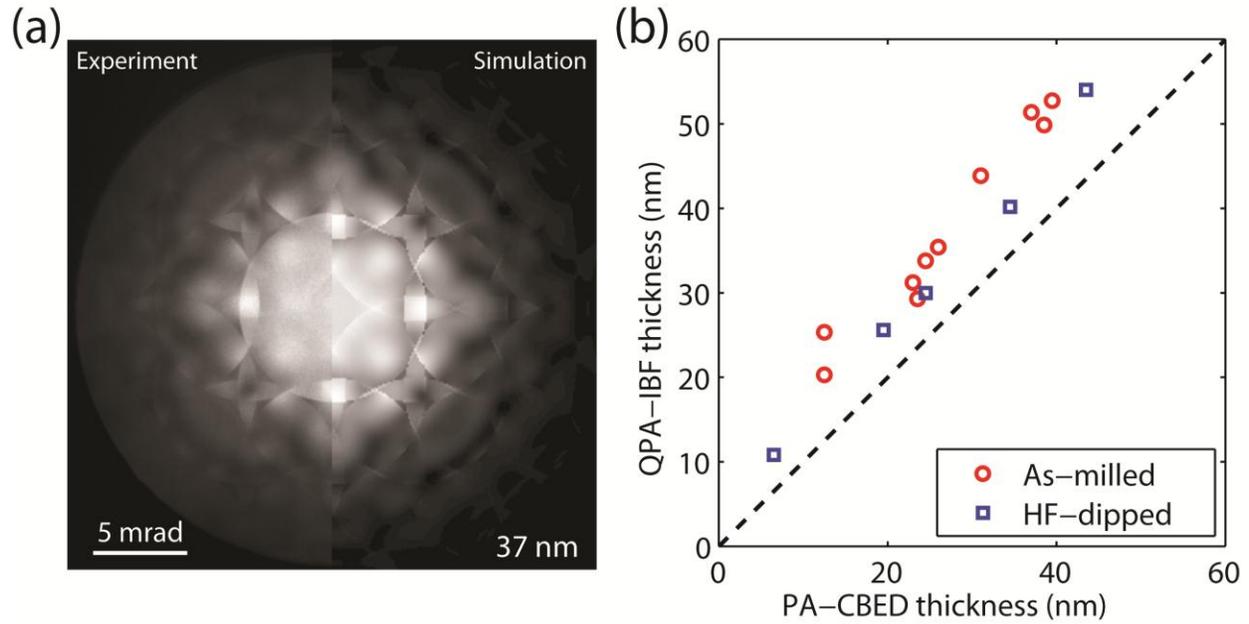

Figure 5. (color online) Comparison of PA-CBED and QPA-IBF methods. (a) The determination of crystalline thickness by matching the PA-CBED with the incoherent CBED simulation. (b) QPA-IBF thickness vs. PA-CBED thickness. The QPA-IBF method gives a thickness measurement higher than of the PA-CBED method due to the presence of amorphous surface layers. In the as-milled sample, the amorphous layer is on average 9.8±1.1 nm (i.e. 4.9 nm/surface). In the HF-dipped sample, the amorphous layer is on average 6.4±1.1 nm (3.2 nm/surface). The HF-dipped sample and the as-milled sample are from two separate STO specimen.

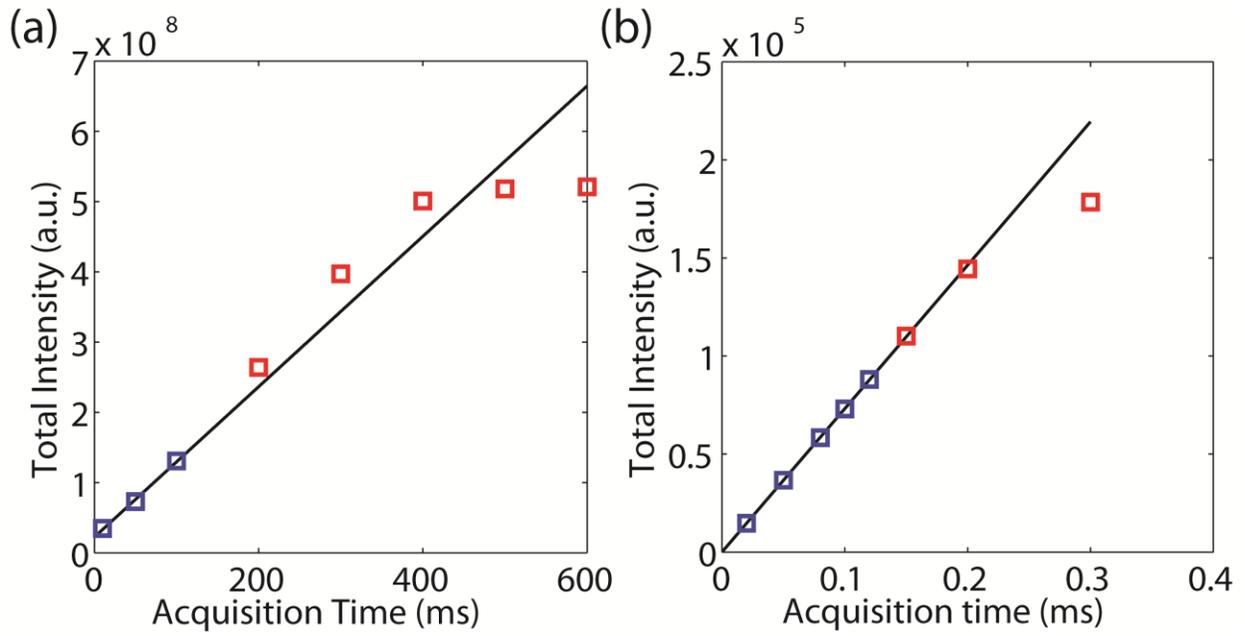

Figure 6. (color online) The comparison of the Ronchigram CCD and the spectrometer CCD. (a) The measurement of the linearity of the Ronchigram CCD. The Ronchigram CCD is highly nonlinear prior to saturation. (b) The measurement of the linearity of the spectrometer CCD. The spectrometer CCD is very linear until saturation.